\documentclass[twocolumn,preprintnumbers,a4paper,
superscriptaddress,floatfix,twoside,prd]{revtex4}

\usepackage{bm}
\usepackage{epsfig}
\usepackage{graphics}
\usepackage{natbib}

\usepackage{fancyh}
\usepackage[l]{floatflt}
\usepackage{epsfig}
\usepackage{amssymb}
\usepackage{latexsym}
\usepackage{times}
\usepackage{slashed}
\usepackage{upgreek}
\usepackage{amsmath}
\usepackage{amsfonts}
\usepackage{amsbsy}
\usepackage{amscd}
\usepackage{bbm}

\newcommand{\Eref}[1]{Eq.~(\ref{#1})}
\newcommand{\Sref}[1]{Sec.~\ref{#1}}
\newcommand{\Fref}[1]{Fig.~\ref{#1}}
\newcommand{\Tref}[1]{Table~\ref{#1}}
\newcommand{\cref}[1]{Ref.~\cite{#1}}

\newcommand{\arxiv}[1]{{\ftn\tt  arXiv:#1}}

\newcommand{\bal}{\begin{align}}
\newcommand{\eal}{\end{align}}
\newcommand{\beqs}{\begin{subequations}}
\newcommand{\eeqs}{\end{subequations}}
\newcommand{\eec}{\end{center}}
\newcommand{\bec}{\begin{center}}

\newcommand{\eem}{\end{matrix}}
\newcommand{\bem}{\begin{matrix}}
\newcommand{\eeq}{\end{equation}}
\newcommand{\beq}{\begin{equation}}
\newcommand{\ba}{\begin{array}}
\newcommand{\ea}{\end{array}}
\newcommand{\bea}{\begin{eqnarray}}
\newcommand{\eea}{\end{eqnarray}}
\newcommand{\baq}{\begin{eqnarray}}
\newcommand{\eaq}{\end{eqnarray}}

\newcommand\eqs[2]{Eqs.~(\ref{#1}) and (\ref{#2})}

\newcommand\eqss[3]{Eqs.~(\ref{#1}), (\ref{#2}) and (\ref{#3})}

\newcommand{\ftn}{\footnotesize}

\newcommand{\etal}{{\it et al.\/}}

\def\to{\rightarrow}

\def\lf{\left(}
\def\rg{\right)}
\newcommand\vev[1]{\langle {#1} \rangle}
\newcommand\vevi[1]{\langle {#1} \rangle_{\rm I}}

\newcommand{\Vhi}{\ensuremath{\widehat V_{\rm I}}}
\newcommand{\Vjhi}{\ensuremath{V_{\rm I}}}
\newcommand{\Hhi}{\ensuremath{\widehat H_{\rm I}}}

\newcommand{\Khi}{\ensuremath{K}}
\newcommand{\Whi}{\ensuremath{W}}
\newcommand{\Vhio}{\ensuremath{\widehat V_{\rm I0}}}
\newcommand{\Ns}{\ensuremath{{N_\star}}}
\newcommand{\ck}{\ensuremath{c_{\rm K}}}

\newcommand{\mP}{\ensuremath{m_{\rm P}}}

\newcommand{\Qef}{\ensuremath{\Lambda_{\rm UV}}}

\def\openone{\leavevmode\hbox{\small1\kern-3.8pt\normalsize1}}
\newcommand{\dV}{\ensuremath{\Delta\widehat V_{\rm I}}}

\newcommand{\fr}{\ensuremath{f_{\cal R}}}
\newcommand{\fk}{\ensuremath{f_{\rm K}}}

\newcommand{\hr}{\ensuremath{F_{\cal R}}}
\newcommand{\hsh}{\ensuremath{F_{\rm sh}}}
\newcommand{\kb}{\ensuremath{K_2}}
\newcommand{\ka}{\ensuremath{K_1}}
\newcommand{\nb}{\ensuremath{N_2}}

\newcommand{\ca}{\ensuremath{c_{\cal R}}}

\newcommand{\ks}{\ensuremath{k_\star}}
\newcommand{\ns}{\ensuremath{n_{\rm s}}}
\newcommand{\as}{\ensuremath{a_{\rm s}}}
\newcommand{\As}{\ensuremath{A_{\rm s}}}

\newcommand{\rcc}{\ensuremath{\mathcal{R}}}
\newcommand{\rce}{\ensuremath{\widehat{\mathcal{R}}}}
\newcommand{\Ve}{\ensuremath{\widehat V}}
\newcommand{\Vf}{\ensuremath{\widehat V_{\rm F}}}

\newcommand{\what}{\ensuremath{\widehat}}

\def\bbet{{\bar\beta}}
\def\al{{\alpha}}

\def\n{\bar{n}}

\def\th{{\theta}}

\newcommand{\sg}{\ensuremath{\phi}}

\newcommand{\sgx}{\ensuremath{\phi_\star}}
\newcommand{\sgf}{\ensuremath{\phi_{\rm f}}}

\newcommand{\ld}{\ensuremath{\lambda}}

\newcommand{\Ld}{\ensuremath{\Lambda}}

\newcommand{\se}{\ensuremath{\widehat \phi}}
\newcommand{\sex}{\ensuremath{\widehat{\phi}_\star}}
\newcommand{\sef}{\ensuremath{\widehat{\phi}_{\rm f}}}
\newcommand{\geu}{\ensuremath{\widehat g}}
\newcommand{\eph}{\ensuremath{\widehat \epsilon}}
\newcommand{\ith}{\ensuremath{\widehat \eta}}

\def\Ka{K\"{a}hler potential}

\def\bcp{{\sc\small Bicep2}/{\it Keck Array}}
\newcommand{\plk}{{\it Planck}}

\def\actc{{\sf\ftn P-ACT-LB-BK18}}

\renewcommand{\refname}{{\bf\scshape References}}

\renewenvironment{subequations}{%
\refstepcounter{equation}%
\setcounter{parentequation}{\value{equation}}%
  \setcounter{equation}{0}
  \ignorespaces
}{%
  \setcounter{equation}{\value{parentequation}}%
  \ignorespacesafterend
}

\begin{document}



\title{\bf\scshape Kinetically Modified Palatini Inflation Meets ACT Data}

\author{\scshape Constantinos Pallis\\ {\it School of Technology,  Aristotle University of
Thessaloniki, Thessaloniki, GR-541 24 GREECE} \\  {\sl e-mail
address: }{\ftn\tt kpallis@auth.gr}}



\begin{abstract}

\noindent {\ftn \bf\scshape Abstract:} We show that the
coexistence of a non-minimal coupling to gravity
$\fr=1+\ca\phi^{n/2}$ with a kinetic mixing of the form
$\fk=\fr^m$ -- where $n=2$ and $4$ and $0.5\leq m\leq10$ --
reconciles chaotic inflation based on the $\phi^n$ potential with
the recent ACT results, if we adopt the Palatini formulation of
gravity. The attainment of inflation allows for subplanckian
inflaton values and energy scales below the cut-off scale of the
corresponding effective theory. The model can be also embedded in
supergravity by introducing two chiral superfields and a monomial
superpotential, linear with respect to the inflaton-accompanying
field. Its stabilization is achieved thanks to a compact
contribution to the \Ka, whose the inflationary part includes an
holomorphic logarithmic term and a real one multiplying a
shift-symmetric quadratic polynomial term.
\\ \\ {\scriptsize {\sf PACs numbers: 98.80.Cq, 04.50.Kd, 12.60.Jv, 04.65.+e}
\hfill {\sl\bfseries Published in} {\sl Phys. Lett. B} {\bf 868},
139739 (2025)

}

\end{abstract}\pagestyle{fancyplain}

\maketitle

\rhead[\fancyplain{}{ \bf \thepage}]{\fancyplain{}{\sl Kinetically
Modified Palatini Inflation Meets ACT Data}}
\lhead[\fancyplain{}{\sl C. Pallis}]{\fancyplain{}{\bf \thepage}}
\cfoot{}

\section{Introduction} \label{intro}

Working in the context of the metric formulation of gravity -- and
in the reduced Planck units with $\mP=M_{\rm P}/\sqrt{8\pi}=1$ --
one can easily verify \cite{old, nmi, nmiq, roest} that the
presence of a non-minimal coupling function
\beq \label{fr} \fr(\sg)=1+\ca\sg^{n/2},  \eeq
between the inflaton $\sg$ and the Ricci scalar $\rcc$, considered
in conjunction with a monomial potential of the type
\beq \label{Vn}\Vjhi(\sg)=\ld^2\sg^n/2^{n/2},\eeq
provides, at the strong $\ca$ limit, an attractor \cite{roest}
towards the (scalar) spectral index
\beq \label{nsnmi} \ns\simeq1-2/\Ns=0.965~~\mbox{for}~~\Ns=55 \eeq
e-foldings with negligible $\ns$ running $\as$ and
tensor-to-scalar ratio $r\simeq0.0036$. Focusing on the most
well-motivated cases -- from Particle Physics point of view --, it
would be emphasized that, in the Palatini formulation
\cite{demir,attrpal} of gravity, the resulting $\ns$ in
\Eref{nsnmi} remains essentially unaltered for $n=4$ whereas it
slightly increases for $n=2$ \cite{linear,attrpal}. Although
perfectly consistent with \bcp\ and \plk\ data \cite{plin}, the
value in \Eref{nsnmi} is in tension with the latest \emph{Data
Release 6} ({\sf\ftn DR6}) from the \emph{Atacama Cosmology
Telescope} ({\sf\ftn ACT}) \cite{act,actin}, combined with the
\emph{cosmic microwave background} ({\sf\ftn CMB} measurements by
\plk\ \cite{plin} and \emph{BICEP/Keck} ({\sf\ftn BK}) \cite{bcp},
together with the \emph{Dark Energy Spectroscopic Instrument}
({\sf\ftn DESI}) {\it Baryon Acoustic Oscillation} ({\sf\ftn BAO})
results \cite{desi}. Indeed, the so-called \actc\ data entails
\cite{actin}
\beq \label{data} \ns=0.974\pm0.0068,~\as  =
0.0062\pm0.0104~~\mbox{and}~~r\leq0.038, \eeq
at 95$\%$ \emph{confidence level} ({\sf\small c.l.}). The
resulting $\ns$ (mainly) significantly affects the viability of
various well-established the last years inflationary models
\cite{plin}. As a consequence, several modifications have been
recently proposed to reconcile with the data, e.g., (non-)minimal
\cite{actlinde,actattr,aoki,maity,rhb,nmact,yin,warm,rhc,rha,act1,act3,act2,act4,act5,act6,oxf}
or Starobinsky \cite{actellis,r2drees,ketov,acttamv,r2a,r2b}
inflation.

Following this avenue, we here focus on a variant of
\emph{non-minimal inflation} ({\sf\ftn nMI}) developed
\cite{nmikr} by introducing a suitable non-canonical kinetic
mixing $\fk(\phi)$ in the inflaton sector. For this reason the
term \emph{kinetically modified nMI} was coined. The key point of
our present approach is the adoption of the Palatini rather than
the metric formulation of gravity -- for a review see
\cref{palreview}. The consequences of this alteration is twofold
\emph{with respect to} ({\ftn\sf w.r.t}) our findings in
\cref{nmikr}: The resulting $\ns$ increases without an
unacceptable elevation of $r$ which can be partially tested in the
near future \cite{det} and the inflationary scale remains below
the \emph{Ultraviolet} ({\sf\ftn UV}) cut-off scale of the
effective theory without the introduction of a new parameter
($\ck$) -- cf. \cref{un1,un2,riotto,sibi,udemir,un4,un3,un0,un5}.
Nonetheless, the kinetic modification is beneficial for Palatini
nMI since it assists to obtain subplanckian values for the
inflaton field. At last, we present for first time, to our
knowledge, a realization of Palatini nMI in the standard
\emph{Supergravity} ({\sf\ftn SUGRA}) following the strategy
introduced in \cref{rube,su11,unvr2}.


\section{Set-up} \label{set}

At the non-SUSY level, our proposal can be formulated in the
\emph{Jordan frame} ({\sf\ftn JF}) where the action of $\phi$ is
given by
\beq \label{action1} {\sf  S} = \int d^4 x \sqrt{-\mathfrak{g}}
\left(-\frac{\fr}{2}\rcc +\frac{\fk}{2}g^{\mu\nu}
\partial_\mu \sg\partial_\nu \sg-
\Vjhi(\sg)\right). \eeq
Here $\mathfrak{g}$ is the determinant of the background
Friedmann-Robertson-Walker metric, $g^{\mu\nu}$ with signature
$(+,-,-,-)$ and we  allow for a kinetic mixing through the
function $\fk(\phi)$ which is conveniently parameterized in terms
of $\fr$ as follows
\beq\label{fk} \fk(\sg)=\fr^m~~\mbox{where}~~ 1/2\leq m\leq10\eeq
is an indicative interval for $m$ which assures an ample variation
of $\fk$. Indeed, any value of $\fk$ can be revealed by the
parametrization of \Eref{fk} selecting conveniently
$m=\ln\fk/\ln\fr$. The considered $m$ margin can be characterized
as natural, since the values are of order unity, and allows us to
overlaid almost the whole observationally favored region, as we
show in \Sref{num}. Contrary to the models in \cref{martin} --
with which ours share the same name -- we do not violate the
quadratic form of the kinetic operator $\dot\sg^2$ for
$\mu=\nu=0$. By performing a conformal transformation \cite{nmi},
according to which we define the \emph{Einstein frame} ({\sf\ftn
EF}) metric $\geu_{\mu\nu}=\fr\,g_{\mu\nu}$, we can write ${\sf
S}$ in the EF as follows
\beq {\sf  S}= \int d^4 x
\sqrt{-\what{\mathfrak{g}}}\left(-\frac12
\rce+\frac12\geu^{\mu\nu} \partial_\mu \se\partial_\nu \se
-\Vhi(\se)\right), \label{action} \eeq
where hat is used to denote quantities defined in the EF. Given
that the metric tensor and the affine connection are treated as
independent variables in the Palatini formalism, the EF
canonically normalized field, $\se$, and potential, $\Vhi$, are
defined as follows
\beq \label{VJe}
\frac{d\se}{d\sg}=J=\frac{\fk^{1/2}}{\fr^{1/2}}=\fr^{\frac{m-1}{2}}~~~\mbox{and}~~~
\Vhi= \frac{\Vjhi}{\fr^2}=\frac{\ld^2\sg^n}{2^{n/2}\fr^{2}}.\eeq
From the expressions in \Eref{VJe} it is clear that $\fk$ through
$m$ determines largely the form of $\se(\sg)$ whereas $\fr$
influences exclusively the curvature of $\Vhi$. Therefore, our
scheme is expected to be observationally more flexible compared to
the traditional (metric or Palatini) nMI where $m=0$.

\section{Inflationary Dynamics} \label{ana}

A period of slow-roll nMI is determined in the EF by the condition
\beq{\ftn\sf
max}\{\eph(\phi),|\ith(\phi)|\}\leq1,\label{srcon}\eeq where the
slow-roll parameters $\eph$ and $\ith$ read
\bea\nonumber\widehat\epsilon&=&\left({\Ve_{\rm
I,\se}\over\sqrt{2}\Ve_{\rm
I}}\right)^2=\frac{n^2}{2\ca\sg^2\fr^{1+m}}~~\mbox{and}~~
\widehat\eta={\Ve_{\rm I,\se\se}\over\Ve_{\rm
I}}=\\
&=&\frac{\eph}{2n^2}\lf4n(n-1)-n(4+n(m+1))\ca\sg^{n/2}\rg,~~\label{sr1}\eea
with the symbol $,\chi$ as subscript denoting derivation w.r.t the
field $\chi$. For the expressions above we employ $J$ in
\Eref{VJe}, without expressing explicitly $\Vhi$ in terms of
$\se$. For some $m$ values, actually, this is not doable due to
the complicate form of the $\se(\sg)$ function which is the
following
\bea  \nonumber \se&=&\sg\fr^{\frac{m+1}{2}}
\;{}_2F_1\lf1,\frac{m}{2}+\frac{2}{n}+\frac12;1+\frac{2}{n};-\ca\sgx^{n/2}\rg
\\ &\simeq& \frac{4\ca^{(m-1)/2}\sg^{1+n(m-1)/4}}{4+n(m-1)}~~\mbox{for}~~\ca\gg1.
 \label{sesg} \eea
Here ${}_2F_1$ is the Gauss hypergeometric function
\cite{wolfram}. Note that our expression reduces to that given in
\cref{attrpal} for $m=0$ whereas for $m=1$ we obtain a case with
$\se=\sg$, recently mentioned for $n=4$ in \cref{un5}.
\Eref{srcon} is saturated at the maximal $\sg$ value, $\sgf$, from
the following two values
\beqs\beq \lf\sg_{1\rm f},\sg_{2\rm
f}\rg\simeq\Bigg(\lf\frac{n^2}{2\ca^{m+1}}\rg^{\frac{2}{4+n(m+1)}},
\lf\frac{n(n+1)}{\ca^{1+m}}\rg^{\frac{2}{4+n(m+1)}}\Bigg)
\label{sgf}\eeq
which satisfy for $\sg\ll1$ the equations
\beq \eph\lf\sg_{1\rm f}\rg=|\ith\lf\sg_{2\rm f}\rg|\simeq1.
\label{srf}\eeq\eeqs

\renewcommand{\arraystretch}{1.1}
\begin{table*}[t!]
\caption{\normalfont  Inflationary predictions for $n=2$ and $4$.}
\begin{tabular}{c|@{\hspace{0.1cm}}c@{\hspace{0.1cm}}|@{\hspace{0.1cm}} c}\toprule
&{$n=2$}& $n=4$\\ \hline
$\ns$ &$1-(m+3)\big(1+1/(2(m+2)\ca^2\Ns)^{1/(m+2)}\big)/(m+2)\Ns$&

$1-(m+2)\big(1+1/(8(m+1)\ca\Ns)^{1/(m+1)}\big)/(m+1)\Ns$\\[0.08cm]
$r$ &$16/(m+2)\Ns\big(2\ca^2(m+2)\Ns)\big)^{1/(m+2)}$&

$16/(m+1)\Ns\big(8\ca(m+1)\Ns)\big)^{1/(m+1)}$\\[0.08cm]
$\as$ &$-(m+3)/(m+2)N^2_\star-{\cal O}\big(N^{-2}_\star
(\ca^2\Ns)^{-1/(m+2)}\big)$&
$-(m+2)/(m+1)N^2_\star-{\cal O}\big(N^{-2}_\star
(\ca\Ns)^{-1/(m+1)}\big)$\\\botrule
\end{tabular}\label{tab2}
\end{table*}

The number of e-foldings $\Ns$ that the scale $\ks=0.05/{\rm Mpc}$
experiences during this nMI and the amplitude $\As$ of the power
spectrum of the curvature perturbations generated by $\sg$ can be
computed using the standard formulae
\begin{equation}
\label{Nhi}  \Ns=\int_{\sef}^{\sex} d\se\frac{\Vhi}{\Ve_{\rm
I,\se}}~~\mbox{and}~~\As^{1/2}= \frac{1}{2\sqrt{3}\, \pi} \;
\frac{\Ve_{\rm I}^{3/2}(\sex)}{|\Ve_{\rm I,\se}(\sex)|},\eeq
where $\sgx~[\sex]$ is the value of $\sg~[\se]$ when $\ks$ crosses
the inflationary horizon. Since $\sgx\gg\sgf$, from \Eref{Nhi} we
find
\bea \nonumber \Ns&=&\frac{\sgx^2}{2n}
\;{}_2F_1\lf-m,4/n;1+4/n;-\ca\sgx^{n/2}\rg\\ &\simeq&
\frac{2\ca^m\sgx^{(nm+4)/2}}{n(nm+4)}~~\mbox{for}~~\ca\gg1.
\label{Nhia}\eea
For $\ca\gg1$, we are able to solve \Eref{Nhia} w.r.t $\sgx$ and
pursue our analytical approach. Indeed, we obtain
\begin{equation}
\label{sgx}\sgx\simeq \lf n\lf nm+4\rg\Ns/2\ca^m \rg^{2/(4+nm)}.
\end{equation}
It is clear that there is a lower bound on $\ca$, above which
$\sgx\leq1$ and so, our proposal can be stabilized against
corrections from quantum gravity. I.e.,
\begin{equation}
\label{cab}\sgx\leq1~~\Rightarrow~~\ca\geq\lf n\lf
nm+4\rg\Ns/2\rg^{1/m}.
\end{equation}
Note that this bound can be imposed only for $m\neq0$, as
advocated in \Sref{intro}, and not within the pure Palatini nMI.
Moreover, the minimal $\ca$ value decreases as $m$ increases and
so problems with the pertubative unitarity -- see below -- become
less acute.

From the rightmost relation in \Eref{Nhi} we can also derive a
constraint between $\ld$ and $\ca$ solving w.r.t $\ld$. I.e.
\beq \label{lan}\ld\simeq2^{\frac{n+4}{4}}(3\As)^\frac12\pi
n\ca^\frac{nm+2}{nm+4}(2/n(nm+4)\Ns)^
{\frac{4+n(m+1)}{2(nm+4)}}.\eeq
%
For $m=0$ we reproduce the $\ld-\ca$ relation obtained in the pure
Palatini nMI which differs slightly from that within the metric
nMI -- cf. \cref{palreview}.

The inflationary observables are found from the relations
\beqs\bea \label{ns} && \ns=\: 1-6\widehat\epsilon_\star\ +\
2\widehat\eta_\star,~~r=16\widehat\epsilon_\star, \\ \label{as} &&
\as =\:2\left(4\widehat\eta_\star^2-(n_{\rm
s}-1)^2\right)/3-2\widehat\xi_\star, \eea\eeqs
where the variables with subscript $\star$ are evaluated at
$\sg=\sgx$ and $\widehat\xi={\Ve_{\rm I,\widehat\phi} \Ve_{\rm
I,\widehat\phi\widehat\phi\widehat\phi}/\Ve^2_{\rm I}}$. For
$\ca\gg1$ we find the following general -- i.e., for any $n$ and
$m$ -- expressions
\beqs\bea\nonumber && \ns=1
-\frac{1}{\Ns}-\frac{n}{(nm+4)\Ns}\nonumber
\\&&
-\lf\frac{2^n}{n^n\ca^4}\rg^\frac{1}{nm+4}\frac{n(m+1)+4}{((nm+4)\Ns)^{1+\frac{n}{nm+4}}};\label{ns1}
\\&&
r=16\lf\frac{2^n
n^{4+n(m-1)}}{\ca^4((nm+4)\Ns)^{4+n(m+1)}}\rg^\frac{1}{nm+4}.
\label{rs1}\eea\eeqs
As regards $\as$, specific expressions for $n=2$ and $4$ are given
in \Tref{tab2} together with the corresponding outputs of
Eqs.~(\ref{ns1}) and (\ref{rs1}). Taking the limit $\ca\to\infty$
for $m=0$, \Eref{ns1} is in agreement with \cref{linear, attrpal}
\beq\label{nsattr} \ns=1 - 1/\Ns - n/4 \Ns,\eeq
if we take into account the appropriate redefinition of $n$.
Therefore, for $n=4$ we arrive again at \Eref{nsnmi} whereas for
$n=2$ we obtain $\ns=1-3/2\Ns$ which reaches \Eref{data} for
$\Ns\simeq60$. The same is true for $n=4$ and $m=1$ -- cf.
\cref{un5}. From the results of \Tref{tab2} we can clearly infer
that increasing $m$, $\ns$ increases whereas $r$ remains
moderately suppressed due to the factor $\ca^{-4/(nm+4)}$.


\section{Effective Cut-Off Scale}\label{uv}

It is well known \cite{udemir} that the absence of an extra term
in the leftmost expression in \Eref{VJe} -- cf. \cref{nmikr} --
deliberates the Palatini nMI for $n=4$ from the problem with the
perturbative unitarity which puzzles the metric nMI \cite{un1,
un2, riotto} -- for another attitude on this issue see
\cref{sibi}. We show here that this nice feature of Palatini nMI
insists even in our cases.

To this end we follow the most restrictive approach analyzing the
small-field behavior of our models in the EF -- cf.
\cref{riotto,un1, un2,nmikr}. We focus on the second term in the
right-hand side of \Eref{action} -- or \Eref{Saction1}, see
\Sref{sugra} below -- for $\mu=\nu=0$ and we expand it about
$\vev{\phi}=0$ in terms of $\se$ -- see \Eref{VJe} -- which
coincides with $\sg$ since $\vev{J}=1$. Our result for some $m$
values can be written as
\beq\nonumber J^2
\dot\phi^2=\dot\sg^2\begin{cases}1&\mbox{for}~~m=1,\\
1+\ca\sg^{n/2}&\mbox{for}~~m=2, \\
\lf1+2\ca\sg^{n/2}+\ca^2\sg^{n}+\cdots\rg&\mbox{for}~~m=3.
\end{cases}\eeq
Similar expressions can be obtained for the other $m$'s too.
Expanding similarly $\Vhi$, see \Eref{VJe}, in terms of $\sg$ we
have
\beq\nonumber
\Vhi=\frac{\ld^2\sg^n}{2^{n/2}}\lf1-2\ca\sg^{\frac{n}{2}}+3\ca^2\sg^n-
4\ca^3\sg^{\frac{3n}{2}}+\cdots\rg \eeq
independently of $m$. If we reinstall $\mP$ in the expressions
above, so that we obtain dimensionless quantities in the bracket,
and introduce $\ca$ in the denominators, we can conclude that the
UV cut-off scale is
\beq \label{qef} \Qef=\mP/\ca^{2/n}~~\mbox{for any $m$.}\eeq
Our result is in accordance with \cref{udemir,un0,un3,un4} for the
$n=4$ and $m=0$ case. Consistency with the effective theory
entails
\beq
\Vhio^{1/4}\leq\Qef~~\mbox{with}~~\Vhio\simeq\ld^2\mP^4/2^{n/2}\ca^2.
\label{vqef}\eeq
If we take into account \Eref{lan}, the condition above yields an
upper bound on $\ca$ for every selected $n$ and $m$. As we show in
\Sref{num}, there is sizable parameter space consistent with this
requirement -- even for the most restrictive case with $n=2$. The
validity of \Eref{vqef} is of crucial importance for the viability
of the effective theory \cite{riotto}, independently from the
hierarchy between $\sgx$ and $\Qef$, since it is widely believed
that dangerous loop corrections depend on the energy scale and not
on the field values. Therefore, the fact that in our models we
obtain $\sgx\gg\Qef$ -- which causes, in principle, concerns
regarding corrections from non-renormalizable terms associated
with $\Qef$ -- does not invalidate our proposal. In addition, a
more elaborated derivation of $\Qef$ \cite{un0} for $n=4$ and
$m=0$ reveals the absence of any relevant problem.

\begin{figure*}[!t]
\includegraphics[width=60mm,angle=-90]{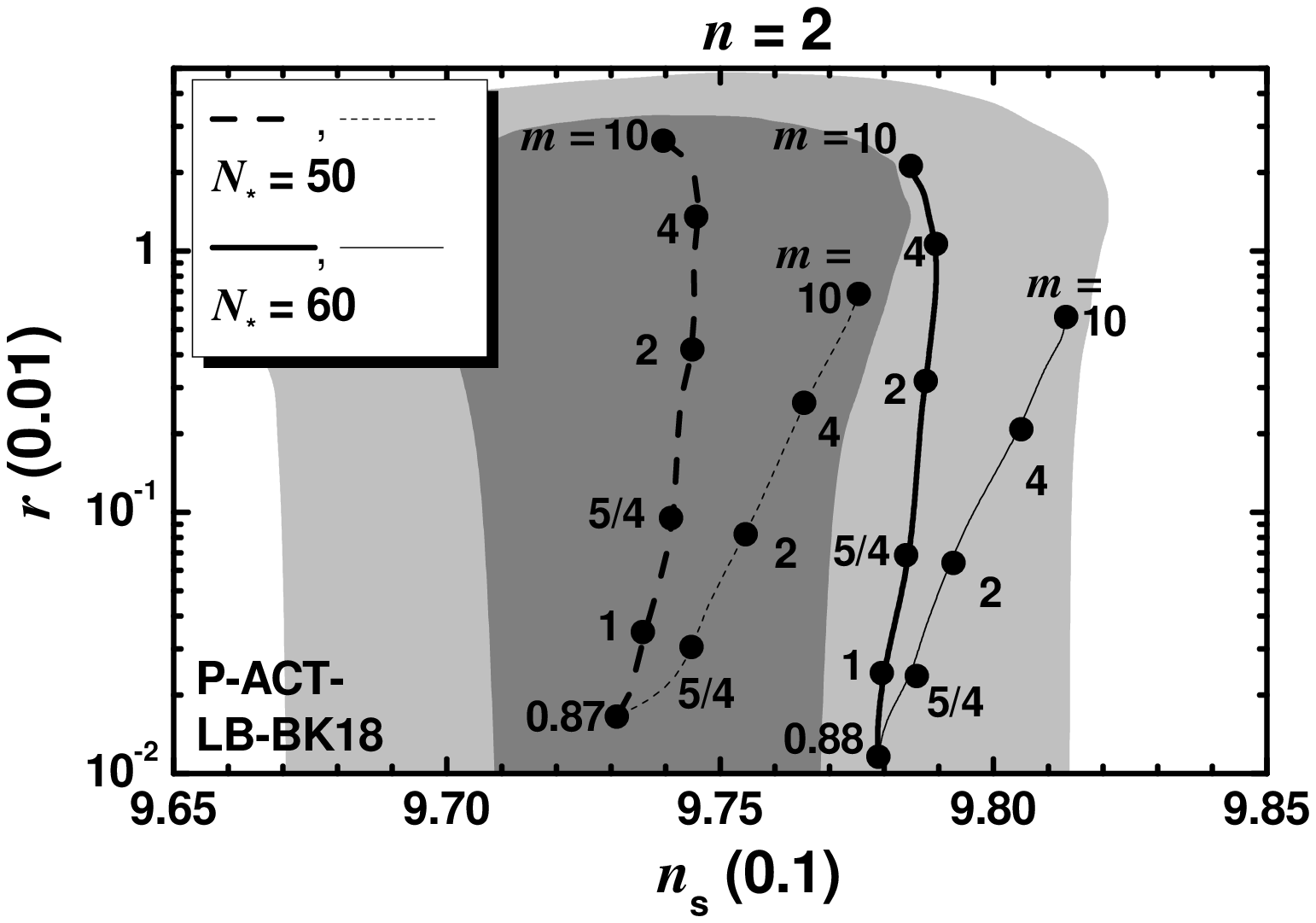}
\includegraphics[width=60mm,angle=-90]{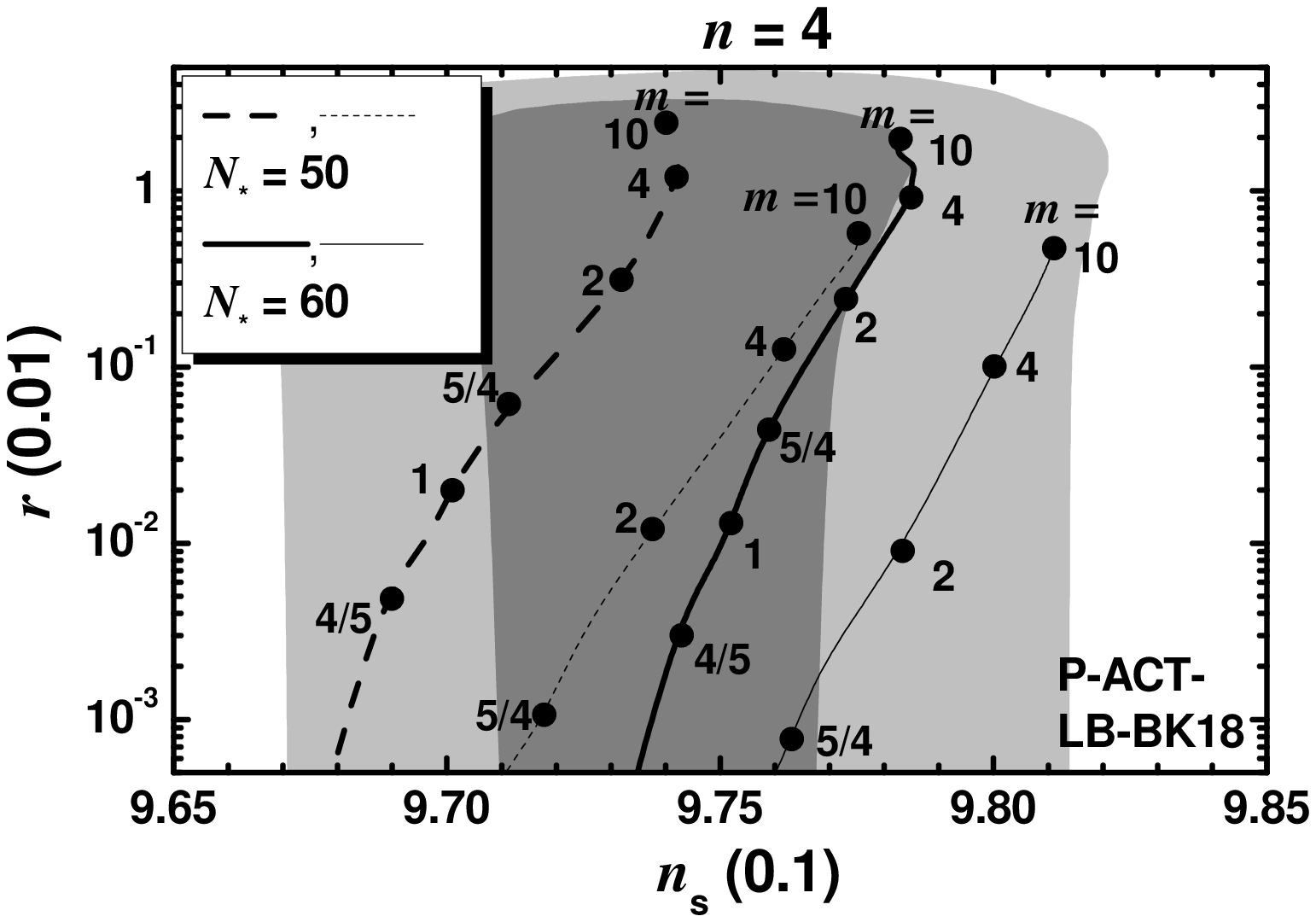}
\caption{\sl Allowed curves in the $\ns-r$ plane for $n=2$ (left
panel) and $4$ (right panel) and various $m$'s indicated on the
curves. We take $\Ns=50$ (dashed lines) or $\Ns=60$ (solid lines)
and minimal (thick lines) or maximal (thin lines) $\ca$ values.
The marginalized joint $68\%$ [$95\%$] c.l. regions from \actc\
data are depicted by the dark [light] shaded
contours.}\label{fig1}
\end{figure*}

\section{Numerical Results}\label{num}

The approximate analytical estimates in \Sref{ana} -- which are
accurate enough for $\ca\gg1$ -- can be verified and extended to
all possible $\ca$ values numerically. Namely, confronting the
quantities in \Eref{Nhi} with the observational requirements
\cite{act}
\begin{equation}
\label{Prob}
50\lesssim\Ns\lesssim60~~\mbox{and}~~\As^{1/2}\simeq4.618\cdot10^{-5},\eeq
we can restrict $\ld$ and $\sgx$ and compute the model predictions
via \eqs{ns}{as}, for any selected $\ca$, $m$ and $n$. Taking the
$\Ns$ range above we assumed that nMI is followed in turn by a
oscillatory phase with mean equation-of-state parameter $w_{\rm
rh}$, radiation and matter domination, with variation \cite{plin,
rehpal,rehpal1,reh1,reh2} of $w_{\rm rh}$ and the reheating
temperature $T_{\rm rh}$ in the following ranges  respectively
\beq 0\lesssim w_{\rm rh}\lesssim1/3~~\mbox{and}~~1~{\rm
TeV}\lesssim T_{\rm rh}\lesssim 1~{\rm EeV}, \label{wtrh}\eeq
with units reinstalled. The $\ca$ values can be further
constrained by the saturation of \eqs{cab}{vqef}, from which the
first one gives the lowest $\ca$ value (for $\sgx=1$) whereas the
second yields the largest one.

The outputs of our numerical approach, encoded as lines in the
$\ns-r$ plane, are compared against the \actc\ data \cite{actin}
in \Fref{fig1} for $n=2$ (left panel) and $n=4$ (right panel). The
variation of $m$ is shown along each line. In each plot we depict
solid lines for $\Ns=60$ and dashed lines for $\Ns=50$. We also
display two thick lines corresponding to the minimal possible
$\ca$ and two thin ones originating from the maximal $\ca$. E.g.,
for $m=1$ and $\Ns=50$ or $60$  we obtain correspondingly
\bea \nonumber 302\lesssim\ca\lesssim 625&\mbox{or}&362\lesssim\ca\lesssim 670~~\mbox{for}~~n=2~~\mbox{and}\\
\nonumber 8\lesssim\frac{\ca}{100}\lesssim
3.5\cdot10^3&\mbox{or}&9.7\lesssim\frac{\ca}{100}\lesssim
4.2\cdot10^3~~\mbox{for}~~n=4.\eea
In the domain of large $\ca$ we assure that the perturbative bound
$\ld=3.5$ is not violated. Decreasing $m$, the needed $\ca$ values
increase and so \Eref{vqef} becomes more relevant. For $n=2$ and
$m=0.88$ and $0.87$ the lower and upper bounds on $\ca$ from
\eqs{cab}{vqef} respectively coincide and so the the two (thick
and thin) dashed and solid lines intersect. On the other hand, the
graph for $n=4$ is extended to lower $m$ values and such an
intersection does not occur for the range of $m$ in \Eref{fk}.
E.g., along the dashed thick lines, with decreasing $m$, we obtain
\beq 0.005\lesssim\Vhio^{1/4}/\Qef\lesssim1~[0.07] ~~\mbox{for
$n=2$ [$n=4$]}. \eeq
On the other hand, as $m$ increases $r$ increases slowly and
slowly due to decrease of $\ca$ -- see \Eref{rs1} -- and so the
upper bound in \Eref{data} is not saturated. Therefore, no
realistic upper bound on $m$ can be found.

Comparing the resulting corridor between the two thick (or thin)
lines in the two panels of \Fref{fig1} we remark that the allowed
region for $n=4$ overlaps better with 1-$\sigma$ range of
\Eref{data}. Indeed, as $n$ increases above $2$ the resulting
corridor moves to the left. E.g., fixing $m=\sgx=1$ we obtain
\beqs \bea0.969\lesssim\ns\lesssim 0.974&\mbox{for}&n=6,\\
0.967\lesssim\ns\lesssim 0.973&\mbox{for}&n=8,\\
0.966\lesssim\ns\lesssim 0.971&\mbox{for}&n=10,\eea\eeqs
with $r$ decreasing as $n$ increases. Therefore, the considered
here $n$ values offer the best prospects for detecting primordial
gravitational waves in future experiments \cite{det}.

Note, finally, that along the curves depicted in \Fref{fig1} the
obtained $\as$ remains negligibly small and therefore it lies
within the margin in \Eref{data}. E.g., along the dashed [solid]
thick lines, it is
$\as\simeq5.3\cdot10^{-4}~[\as\simeq3.6\cdot10^{-4}]$ for $n=2$
whereas for $n=4$, we obtain
\beq
5.2\lesssim-\as/10^{-4}\lesssim6.3~[3.7\lesssim-\as/10^{-4}\lesssim4.3]
\eeq
with decreasing $m$. These results are of the same order of
magnitude with those obtained within conventional (metric or
Palatini) nMI.

\section{Supergravity Embedding}\label{sugra}

Starting from the EF action in SUGRA we attempt to reproduce the
ingredients of our model in \Eref{VJe} without caring if the
complete JF SUGRA action is consistent with \Eref{action1}. This
task is not realized up to now, to our knowledge,  for the
Palatini formalism.

Adopting the strategy of \cref{rube, su11, unvr2} we consider two
gauge-singlet chiral superfields, i.e., $z^\al=\Phi, S$, with
$\Phi$ ($\al=1$) and $S$ ($\al=2)$ being the inflaton and a
``stabilizer'' field respectively. The EF action for $z^\al$'s
within SUGRA can be written as
\beqs \beq\label{Saction1}  \hspace*{-2.mm}{\sf S}=\int d^4x
\sqrt{-\what{ \mathfrak{g}}}\lf-\frac{1}{2}\rce +K_{\al\bbet}
\geu^{\mu\nu}\partial_\mu z^\al \partial_\nu z^{*\bbet}-\Vf\rg
\eeq
where summation is taken over the scalar fields $z^\al$, star
($^*$) denotes complex conjugation, $\Khi$ is the \Ka\ with
$K_{\al\bbet}=K_{,z^\al z^{*\bbet}}$ and
$K^{\al\bbet}K_{\bbet\gamma}=\delta^\al_{\gamma}$. Also $\Vf$ is
the EF F--term SUGRA potential given by
\beq \Vf=e^{\Khi}\left(K^{\al\bbet}(D_\al W) (D^*_\bbet
W^*)-3{\vert W\vert^2}\right),\label{Vsugra} \eeq \eeqs
where $D_\al W=W_{,z^\al} +K_{,z^\al}W$ with $\Whi$ being the
superpotential. Therefore, the desired SUGRA embedding of our
models requires the determination of $K$ and $W$.

Our task can be facilitated, if we express $\Phi$ and $S$
according to the parametrization
\beq \Phi=\:{\phi\,e^{i
\th}}/{\sqrt{2}}\>\>\>\mbox{and}\>\>\>S=\:(s +i\bar
s)/\sqrt{2}\,,\label{cannor} \eeq
and determine the inflationary track by the constraints
\beq \label{inftr}
\vevi{S}=\vevi{\Phi-\Phi^*}=0,~\mbox{or}~~\vevi{s}=\vevi{\bar
s}=\vevi{\th}=0,\eeq
where the symbol $\vevi{Q}$ stands for the value of a quantity $Q$
during nMI. $\Vjhi$ in \Eref{Vn} can be produced, in the flat
limit, by
\beq \label{Wn} W=\ld
S\Phi^{n/2}~~\mbox{since}~~\vevi{|W_{,S}|^2}=\Vjhi.\eeq
The form of $W$ can be uniquely determined if we impose an $R$
symmetry, under which $S$ and $\Phi$ have charges $1$ and $0$, and
a global $U(1)$ symmetry with assigned charges to $S$ and $\Phi$
$-1$ and $2/n$. The latter is violated, though, in the proposed
$K$ which is judiciously chosen so that $J$ and $\Vhi$ in
\Eref{VJe} are reproduced. More specifically, $\Vhi$ has to be
derived from the only surviving term in \Eref{Vsugra} along the
track in \Eref{inftr}, which is
\beq \label{1Vhio}\vevi{\Vf}=\vevi{e^{K}K^{SS^*}\,
|W_{,S}|^2}\,.\eeq

The proposed $K$ includes two contributions without mixing between
$\Phi$ and $S$, i.e.,
\beq K=\ka+\kb, \label{ktot}\eeq
from which $\kb$ successfully stabilizes $S$ along the path in
\Eref{inftr} without invoking higher order terms -- cf.
\cref{nmikr}. We adopt the form \cite{su11}
\beq \kb=\nb\ln\lf1+{|S|^2/\nb}\rg~~\mbox{with}~~0<\nb<6
\label{kb}\eeq
which parameterizes \cite{su11} the compact manifold $SU(2)/U(1)$
with curvature $2/\nb$. On the other hand, $\ka$ depends on $\Phi$
and has to incorporate $\fr$ and $\fk$ in \eqs{fr}{fk}. To this
end we introduce the functions
\beq \label{hr}
\hr(\Phi)=1+2^{\frac{n}{4}}\ca\Phi^{\frac{n}{2}}~~\mbox{and}~~\hsh=-\frac12(\Phi-\Phi^*)^2,
\eeq
from which \hr\ is holomorphic, reducing to $\fr$ along the path
in \Eref{inftr}, whereas $\hsh$ is real, assisting us to
incorporate the non-canonical kinetic mixing in \Eref{VJe} -- cf.
\cref{nmikr}. In particular,
\beq \vevi{\hr}=\fr,~\vevi{F_{\rm sh}}=0~~\mbox{with}~~F_{\rm
sh,\Phi\Phi^*}=1. \label{frsh}\eeq
Therefore, \hr\ can be used to derive the denominator of \Vhi\ in
\Eref{VJe} whereas the shift-symmetric $\hsh$, multiplied by a
real function of $\Phi$, lets intact $\vevi{\Vf}$ but it may
contribute to the normalization of $\phi$. The relevant functions
$\fk$ and $J$ can be found in terms of $\hr$ during nMI as follows
\beq \fk=\vevi{(\hr+\hr^*)/2}^m~~\mbox{and}~~
J^2=\vevi{(\hr+\hr^*)/2}^{m-1}.\label{vevikj}\eeq
Employing the ingredients above, a simple, suitable form of $\ka$
in \Eref{ktot} is
\beq \ka=-\ln\hr-\ln\hr^*+\lf\frac{\hr+\hr^*}{2}\rg^{m-1}\hsh.
\label{ka}\eeq
It is worth noticing that contrary to the embedding of metric nMI
in SUGRA -- cf. \cref{nmikr} --, $\hr$ and $\hr^*$ enter two
different terms in the right-hand side of \Eref{ka} and so they
give zero contribution into $K_{\Phi\Phi^*}$. Note that for $m=1$,
$\hsh$ and $\hr$ are totally decoupled. If, in addition, we take
$n=4$ then $K$ is totaly quadratic. Therefore, this is the most
well-motivated case from the point of view of SUGRA. Mild (of
order $0.1$) variation of the prefactors of the logarithms is
expected to have some impact on the observational results -- cf.
\cref{nmiq,jhep} -- at the cost of some tuning, though, that we
avoid in our current investigation.

The appropriateness of $K$ in \eqss{ktot}{kb}{ka} can be verified
if we insert it into \Eref{1Vhio} and prove
\beq \vevi{K_{\Phi\Phi^*}}= J^2~~\mbox{and}~~\vevi{\Vf}=\Vhi,
\label{ver}\eeq
where $J$ and $\Vhi$ are given in \Eref{VJe}. Indeed, the first of
the expressions above is derived if we take into account that
\beqs\beq\vevi{K_{\Phi\Phi^*}}=\vevi{K_{1\Phi\Phi^*}}=J^2\vevi{F_{\rm
sh,\Phi\Phi^*}}=J^2\label{vevik}\eeq
-- see \Eref{frsh} -- whereas the second one, if we consider
\Eref{Wn} and notice that
\beq
\vevi{e^K}=\fr^{-2}~~\mbox{and}~~\vevi{K_{SS^*}}=\vevi{K^{SS^*}}=1.\label{ekss}\eeq
\eeqs
Consequently, both expressions in \Eref{VJe} can be recovered from
the proposed $W$ and $K$ in \eqs{Wn}{ktot} via \Eref{ver}.

\begin{table}[t!]
\caption{\normalfont Mass spectrum along the path in
\Eref{inftr}.}
\begin{ruledtabular}
\begin{tabular}{c|c|c}
%
{\sc Fields}&{\sc Eingestates} & {\sc Masses Squared}\\\hline
\multicolumn{3}{c}{Scalars}\\ \hline
$1$ real &$\what \th$ & $(6+3n^2\ca/2\sg^{2-n/2}\fr^{m+1})\Hhi^2$\\
$2$ real&$\what{s},\what{\bar s}$ & $(6/\nb+3n^2/2\sg^{2}\fr^{m+1})\Hhi^2$\\
\hline%
\multicolumn{3}{c}{Spinors}\\\hline
$2$ Weyl &$({\what{\psi}_{S}\pm \what{\psi}_{\Phi})/\sqrt{2}}~~$&
$3n^2\Hhi^2/2\sg^{2}\fr^{m+1}$
\end{tabular}
\end{ruledtabular}\label{tab1}
\end{table}
%

To consolidate the SUGRA embedding of our models we further verify
that the configuration in \Eref{inftr} is stable w.r.t the
excitations of the non-inflaton fields which are canonically
normalized fields by the relations
\beq  \label{Jg} \what{\th}= J\th\sg~~\mbox{and}~~(\what
s,\what{\bar s})={(s,\bar s)}.\eeq
In particular, we find the expressions of the masses squared
$\what m^2_{\chi^\al}$ (with $\chi^\al=\th$ and $s$) arranged in
\Tref{tab1}. These expressions assist us to appreciate the role of
$\nb$ with $0<\nb<6$ in retaining positive $\what m^2_{s}$ -- in
practise we use $\nb=1$. Also we confirm that $\what
m^2_{\chi^\al}\gg\Hhi^2=\Vhio/3$ for $\sgf\leq\sg\leq\sgx$. In
\Tref{tab1} we display the masses $\what m^2_{\psi^\pm}$ of the
corresponding fermions too. We define
$\what\psi_{S}=\sqrt{K_{SS^*}}\psi_{S}$ and
$\what\psi_{\Phi}=\sqrt{K_{\Phi\Phi^*}}\psi_{\Phi}$ where
$\psi_\Phi$ and $\psi_S$ are the Weyl spinors associated with $S$
and $\Phi$ respectively.

Inserting the derived mass spectrum in the well-known
Coleman-Weinberg formula, we  can find the one-loop radiative
corrections, $\dV$ to $\Vhi$ -- cf. \cref{nmiq, jhep}. It can be
verified that our results are immune from $\dV$, provided that the
renormalization group mass scale $\Lambda_{\rm CW}$, is determined
by requiring $\dV(\sgx)=0$ or $\dV(\sgf)=0$. E.g., imposing the
first of these conditions we find  $\Ld_{\rm
CW}\simeq(3.2-41)\cdot10^{-6}$ for $n=2$ and $\Ld_{\rm
CW}\simeq(7.75-39)\cdot10^{-6}$ for $n=4$ along the thick dashed
lines of \Fref{fig1}. Under these circumstances, our results in
the SUGRA set-up can be reproduced by using exclusively the
ingredients of \Eref{VJe} as in the non-SUSY set-up.

\section{Conclusions}

Prompted by the \actc\ data \cite{actin} which is just marginally
compatible with the conventional (metric or Palatini) nMI
\cite{nmi,roest,attrpal}, we proposed a variant within Palatini
gravity which assures comfortable consistency with the current
observations.

Namely, working along the lines of \cref{nmikr}, we considered a
non-canonical kinetic mixing, \Eref{fk}, -- involving the exponent
$m$ -- in the inflaton sector,  apart from the non-minimal
coupling to gravity, \Eref{fr}, which is associated with the
potential in \Eref{Vn}. Confining $m$ to the range
$(0.87~[0.5]-10)$ for $n=2$ [$n=4$] we achieved observational
predictions which offer a nice covering of the present data on
$\ns$ -- its compatibility especially with the $n=4$ case for low
$m$ values is really impressive, as shown in \Fref{fig1}. Part of
the resulting $r$ values there could be accessible in the near
future. Our solutions can be attained with subplanckian values of
the inflaton, requiring large $\ca$'s, and without causing serious
problems with the perturbative unitarity since the
unitarity-violation scale remains well above the inflationary one.
It is gratifying, in addition, that a sizable fraction of the
allowed parameter space of our models (with $\ca\gg1$) can be
studied analytically.

Our setting can be elegantly implemented in SUGRA too, employing
the super- and \Ka s given in \eqs{Wn}{ktot} with individual
contributions in \eqs{kb}{ka}. The non-minimal coupling of the
inflaton to gravity is generated by an holomorphic logarithmic
contribution into $K$ whereas its kinetic mixing is expressed by a
real function multiplying a shift-symmetric quadratic polynomial
term. Our construction can be extended for a gauge non-singlet
Higgs superfield as done in \cref{nmhkr,jhep} for the metric
formulation.

One possible shortcoming of our proposal is that it employs one
extra parameter ($m$) compared to the original model of nMI
reducing, thereby, its predictability. As another consequence,
however, the enhancement of $\ns$ for $n=2$ with $m>0$ may be more
drastic than that presented in \cref{linear,actlinde}. Moreover,
within our scheme, the reheating phase is not constrained as,
e.g., in \cref{rha,rhc, act5, r2drees,rhb} and corrections to the
inflationary potential from other sectors of the theory are not
required as, e.g., in \cref{act6,oxf,warm}.



\paragraph*{\small\bfseries\scshape Acknowledgments} {\small I would like to thank
S. Ketov for interesting discussions.}


\def\ijmp#1#2#3{{\sl Int. Jour. Mod. Phys.}
{\bf #1},~#3~(#2)}
\def\plb#1#2#3{{\sl Phys. Lett. B }{\bf #1}, #3 (#2)}
\def\prl#1#2#3{{\sl Phys. Rev. Lett.}
{\bf #1},~#3~(#2)}
\def\rmp#1#2#3{{Rev. Mod. Phys.}
{\bf #1},~#3~(#2)}
\def\prep#1#2#3{{\sl Phys. Rep. }{\bf #1}, #3 (#2)}
\def\prd#1#2#3{{\sl Phys. Rev. D }{\bf #1}, #3 (#2)}
\def\npb#1#2#3{{\sl Nucl. Phys. }{\bf B#1}, #3 (#2)}
\def\npps#1#2#3{{Nucl. Phys. B (Proc. Sup.)}
{\bf #1},~#3~(#2)}
\def\mpl#1#2#3{{Mod. Phys. Lett.}
{\bf #1},~#3~(#2)}
\def\jetp#1#2#3{{JETP Lett. }{\bf #1}, #3 (#2)}
\def\app#1#2#3{{Acta Phys. Polon.}
{\bf #1},~#3~(#2)}
\def\ptp#1#2#3{{Prog. Theor. Phys.}
{\bf #1},~#3~(#2)}
\def\n#1#2#3{{Nature }{\bf #1},~#3~(#2)}
\def\apj#1#2#3{{Astrophys. J.}
{\bf #1},~#3~(#2)}
\def\mnras#1#2#3{{MNRAS }{\bf #1},~#3~(#2)}
\def\grg#1#2#3{{Gen. Rel. Grav.}
{\bf #1},~#3~(#2)}
\def\s#1#2#3{{Science }{\bf #1},~#3~(#2)}
\def\ibid#1#2#3{{\it ibid. }{\bf #1},~#3~(#2)}
\def\cpc#1#2#3{{Comput. Phys. Commun.}
{\bf #1},~#3~(#2)}
\def\astp#1#2#3{{Astropart. Phys.}
{\bf #1},~#3~(#2)}
\def\epjc#1#2#3{{Eur. Phys. J. C}
{\bf #1},~#3~(#2)}
\def\jhep#1#2#3{{\sl J. High Energy Phys.}
{\bf #1}, #3 (#2)}
\newcommand\jcap[3]{{\sl J.\ Cosmol.\ Astropart.\ Phys.\ }{\bf #1}, #3 (#2)}
\newcommand\jcapn[4]{{\sl J.\ Cosmol.\ Astropart.\ Phys.\ }{\bf #1}, #3, no.~#4 (#2)}
\newcommand\njp[3]{{\sl New.\ J.\ Phys.\ }{\bf #1}, #3 (#2)}


\begin{thebibliography}{99}
 \section*{\refname}  

\bibitem{old}  D.S. Salopek, J.R. Bond and J.M.
Bardeen, {\it Designing Density Fluctuation Spectra in Inflation,}
{\sl Phys. Rev. D }{\bf 40}, 1753 (1989).


\bibitem{nmi} C. Pallis, \textit{Non-Minimally gravity-Coupled Inflationary Models},
\plb{692}{2010}{287} [\arxiv{1002.4765}].


\bibitem{nmiq} C.~Pallis and Q.~Shafi, \textit{Gravity Waves From Non-Minimal
Quadratic Inflation}, {\sl J. Cosmol. Astropart. Phys.}
\textbf{03}, {023} (2015) [\texttt{arXiv:1412.3757}]

\bibitem{roest} R. Kallosh, A. Linde and D. Roest, {\it Universal Attractor for Inflation at Strong
Coupling}, {\sl Phys. Rev. Lett.} {\bf 112}, 011 303 (2014)
[\arxiv{1310.3950}].


\bibitem{demir} F. Bauer and D. A. Demir,
{\it Inflation with Non-Minimal Coupling: Metric versus Palatini
Formulations}, {\it Phys. Lett. B} {\bf 665}, 222 (2008)
[\arxiv{0803.2664}].


\bibitem{attrpal} L.~J\"arv, A.~Racioppi and T.~Tenkanen,
{\it Palatini side of inflationary attractors,} {\sl Phys. Rev. D}
\textbf{97}, no.~8, 083513 (2018) [\arxiv{1712.08471}].


\bibitem{linear} C. Dioguardi and A. Karam, {\it Palatini Linear Attractors Are Back in
ACTion,} {\sl Phys. Rev. D }\textbf{111}, no.~12, 123521 (2025)
[\arxiv{2504.12937}].


\bibitem{plin} Y.~Akrami {\it et al.} [\plk\ Collaboration], {\it Planck
2018 results. X. Constraints on inflation}, {\sl Astron.
Astrophys. }\textbf{641}, A10 (2020) [\arxiv{1807.06211}].


\bibitem{act} T.~Louis \textit{et al.} [ACT Collaboration],
{\it The Atacama Cosmology Telescope: DR6 Power Spectra,
Likelihoods and $\Lambda$CDM Parameters}, \arxiv{2503.14452}.

\bibitem{actin} E.~Calabrese \textit{et al.} [ACT  Collaboration],
{\it The Atacama Cosmology Telescope: DR6 Constraints on Extended
Cosmological Models}, \arxiv{2503.14454}.

\bibitem{desi} A.G. Adame \etal\ [DESI collaboration], {\it DESI 2024 VI: cosmological constraints from
the measurements of baryon acoustic oscillations},
\jcap{02}{2025}{021} [\arxiv{2404.03002}].

\bibitem{bcp} P.A.R. Ade \etal\ [BICEP, Keck collaboration], {\it Improved
Constraints on Primordial Gravitational Waves using Planck, WMAP,
and BICEP/Keck Observations through the 2018 Observing Season},
{\sl Phys. Rev. Lett. }{\bf 127}, 151301  (2021) [\arxiv{
2110.00483}].


\bibitem{actlinde} R. Kallosh, A. Linde and D. Roest, {\it A simple scenario for the
last ACT}, \arxiv{2503.21030}.


\bibitem{aoki} S. Aoki, H. Otsuka and R. Yanagita, {\it Higgs-Modular
Inflation}, \arxiv{2504.01622}.

\bibitem{rhc} R.~Mondal, S.~Mondal and A.~Chakraborty, {\it Constraining
Reheating Temperature, Inflaton-SM Coupling and Dark Matter Mass
in Light of ACT DR6 Observations,} \arxiv{2505.13387}.

\bibitem{rhb} L. Liu, Z. Yi, and Y. Gong, {\it Reconciling Higgs Inflation with ACT
Observations through Reheating}, \arxiv{2505.02407}.

\bibitem{rha} S.~Maity,
{\it ACT-ing on inflation: Implications of non Bunch-Davies
initial condition and reheating on single-field slow roll models,}
\arxiv{2505.10534}.

\bibitem{act5} M.R.~Haque, S.~Pal and D.~Paul, {\it ACT DR6 Insights on the
Inflationary Attractor models and Reheating,} \arxiv{2505.01517}.


\bibitem{nmact}  Q. Gao, Y. Gong, Z. Yi and F. Zhang, {\it Non-minimal coupling in
light of ACT,} \arxiv{2504.15218}.


\bibitem{maity} M.R.~Haque and D.~Maity, {\it Minimal Plateau Inflation in light
of ACT DR6 Observations,} \arxiv{2505.18267}.


\bibitem{actattr} C. Dioguardi, A.J. Iovino and A. Racioppi, {\it Fractional
attractors in light of the latest ACT observations}, \arxiv{2504.
02809}.

\bibitem{act1} J.~McDonald, {\it Higgs Inflation with Vector-Like Quark
Stabilisation and the ACT spectral index,} \arxiv{2505.07488}.

\bibitem{yin}  W.~Yin, {\it Higgs-like inflation under ACTivated mass,}
\arxiv{ 2505.03004}

\bibitem{act2} Z.~Yi, X.~Wang, Q.~Gao and Y.~Gong, {\it Potential Reconstruction
from ACT Observations Leading to Polynomial $\alpha$-Attractor,}
\arxiv{2505.10268}.

\bibitem{act3} Z.Z.~Peng, Z.C.~Chen and L.~Liu, {\it The polynomial potential
inflation in light of ACT observations,} \arxiv{2505.12816}.

\bibitem{act4}  M.~He, M.~Hong and K.~Mukaida,
{\it Increase of $n_s$ in regularized pole inflation \&
Einstein-Cartan gravity,} \arxiv{2504.16069}.


\bibitem{act6} I.D.~Gialamas, A.~Karam, A.~Racioppi and M.~Raidal,
{\it Has ACT measured radiative corrections to the tree-level
Higgs-like inflation?,} \arxiv{2504.06002}.

\bibitem{oxf} W.J.~Wolf,
{\it Inflationary attractors and radiative corrections in light of
ACT,} \arxiv{2506.12436}.

\bibitem{warm} A. Berera, S. Brahma, Z. Qiu, R.O. Ramos and G.S. Rodrigues,
{\it The early universe is ACT-ing warm}, {\it Phys. Rev. D }{\bf
111}, 123527 (2025) [\arxiv{2504.02655}].

\bibitem{actellis} I. Antoniadis, J. Ellis, W. Ke, D.V. Nanopoulos and K.A. Olive,
{\it How Accidental was Inflation?}, \arxiv{2504.12283}.

\bibitem{acttamv} I.D. Gialamas, T. Katsoulas and K. Tamvakis, {\it Keeping the
relation between the Starobinsky model and no-scale supergravity
ACTive}, \arxiv{2505.03608}.

\bibitem{ketov} A. Addazi,  Y. Aldabergenov and S.V. Ketov, {\it Curvature
corrections to Starobinsky inflation can explain the ACT results},
\arxiv{2505.10305}.

\bibitem{r2a} Yogesh, A. Mohammadi, Q. Wu and
T. Zhu, {\it Starobinsky like inflation and EGB gravity in the
light of ACT}, \arxiv{ 2505.05363}.

\bibitem{r2b} M.R. Haque, S. Pal and D. Paul, {\it Improved Predictions on
Higgs-Starobinsky Inflation and Reheating with ACT DR6 and
Primordial Gravitational Waves}, \arxiv{2505.04615}.

\bibitem{r2drees} M. Drees and Y. Xu, {\it Refined Predictions for Starobinsky Inflation
and Post-inflationary Constraints in Light of ACT}, {\sl Phys.
Lett. B }\textbf{867}, 139612 (2025) [\arxiv{2504.20757}].

\bibitem{nmikr} C.~Pallis, \textit{Kinetically Modified Non-Minimal Chaotic
Inflation}, {\sl Phys. Rev. D }\textbf{91}, no.~12, 123508 (2015)
[\arxiv{1503.05887}].

\bibitem{palreview} T.~Tenkanen,
{\it Tracing the high energy theory of gravity: an introduction to
Palatini inflation}, {\sl Gen. Rel. Grav. }\textbf{52}, no.4, 33
(2020) [\arxiv{2001.10135}].

\bibitem{det} P.~Creminelli \etal\ {\it Detecting Primordial $B$-Modes after Planck,}
\jcap{11}{2015}{031} [\arxiv{ 1502.01983}].

\bibitem{udemir} F. Bauer and D.A. Demir, {\sl Higgs-Palatini Inflation and Unitarity},
{\sl Phys. Lett. B} {\bf 698}, 425 (2011) [\arxiv{1012.2900}].

\bibitem{un1} L.F. Barbon and J.R. Espinosa, {\it On the Naturalness of Higgs
Inflation}, \prd{79}{2009}{081302} [\arxiv{0903.0355}].

\bibitem{un2}  C.P. Burgess, H.M. Lee and M. Trott, {\it Comment on Higgs Inflation
and Naturalness}, \jhep{07}{2010}{007} [\arxiv{1002.2730}].


\bibitem{riotto} A.~Kehagias, A.M.~Dizgah and A.~Riotto, {\it Remarks on the
Starobinsky model of inflation and its descendants},
\prd{89}{2014}{043527} [\arxiv{1312.1155}].


\bibitem{un0} J.~McDonald, {\it Does Palatini Higgs Inflation Conserve
Unitarity?}, \jcap{04}{2021}{069} [\arxiv{ 2007.04111}].

\bibitem{un3} I. Antoniadis, A. Guillen and K. Tamvakis,
{\it Ultraviolet behaviour of Higgs inflation models},
\jhep{08}{2021}{018} [\arxiv{2106.09390}]; {\it Addendum to
``Ultraviolet behaviour of Higgs inflation models''},
\jhep{05}{2022}{074} [\arxiv{2203.10040}].


\bibitem{un4}  A. Ito, W. Khaterd and S. Rasanen, {\it Tree-level unitarity
in Higgs inflation in the metric and the Palatini formulation},
\jhep{06}{2022}{164} [{\tt\ftn arXiv:2111.05621}].

\bibitem{un5} T.~Steingasser, M.P.~Hertzberg and D.I.~Kaiser, {\it Precision
Unitarity Calculations in Inflationary Models,} {\tt\ftn arXiv:
2505.20386}.



\bibitem{sibi}  F. Bezrukov, A. Magnin, M. Shaposhnikov and
S. Sibiryakov, {\it Higgs inflation: consistency and
generalisations}, \jhep{01}{2011}{016} [\arxiv{1008.5157}].



\bibitem{rube}  R.~Kallosh, A.~Linde and T.~Rube, {\it General inflaton potentials in supergravity,}
\prd{83}{2011}{043507} [\arxiv{1011.5945}].

\bibitem{su11} C.~Pallis and N.~Toumbas, \textit{Starobinsky-Type Inflation With Products of K\"ahler
Manifolds}, \jcap{05}{2016}{no. 05, 015} [\arxiv{1512.05657}].

\bibitem{unvr2} C.~Pallis,
\textit{Starobinsky Inflation with T-Model K\"ahler Geometries},
{\sl Universe} {\bf 11}, no.~3, 75 (2025) [\arxiv{2502.00636}].

\bibitem{martin} L.~Lorenz, J.~Martin and C.~Ringeval, {\it Constraints on
Kinetically Modified Inflation from WMAP5,} {\sl Phys. Rev. D}
\textbf{78}, 063543 (2008) [\arxiv{0807.2414}].

\bibitem{wolfram} {\ftn\sf http://functions.wolfram.com/HypergeometricFunctions}.


\bibitem{rehpal} T. Takahashi and T. Tenkanen, {\it Towards distinguishing
variants of non-minimal inflation}, \jcap{04}{2019}{035}
[\arxiv{1812.08492}].

\bibitem{rehpal1} D.Y.~Cheong, S.M.~Lee and S.C.~Park,
{\it Reheating in models with non-minimal coupling in metric and
Palatini formalisms,} \jcapn{02}{2022}{029}{02}
[\arxiv{2111.00825}].

\bibitem{reh1} L. Dai, M. Kamionkowski and J. Wang, {\it Reheating constraints to
inflationary models,} \prl{113}{2014}{041302} [\arxiv{1404.6704}].

\bibitem{reh2} J.L. Cook, E. Dimastrogiovanni, D.A. Easson and L.M. Krauss,
{\it Reheating predictions in single field inflation},
\jcap{04}{2015}{047} [\arxiv{1502.04673}].


\bibitem{jhep} G.~Lazarides and C.~Pallis, \textit{Shift Symmetry and Higgs Inflation in Supergravity with
Observable Gravitational Waves}, {\sl J. High Energy Phys.} {\bf
11}, 114 (2015) [\arxiv{1508.06682}].

\bibitem{nmhkr} C.~Pallis, \textit{Kinetically Modified Non-Minimal Higgs
Inflation in Supergravity}, {\sl Phys. Rev. D} {\bf 92}, no. 12,
121305(R) (2015) [\arxiv{1511.01456}].

\end{thebibliography}
\end{document}